\documentclass[aps,prb,superscriptaddress,amsmath,amssymb,amsfonts,twocolumn]{revtex4}

\DeclareMathOperator{\sgn}{sgn}

\newcommand{\bracket}[1]{\langle #1 \rangle}

\newcommand{\ek}{\epsilon_{\vec{k}}}
\usepackage{graphicx}
\usepackage{bm}
\usepackage{bbold}
\newcommand{\svec}[1]{\bm{#1}}
\renewcommand{\vec}[1]{\mathbf{#1}}

\usepackage{color}

\begin{document}

\title{Mean Field study of the heavy fermion metamagnetic transition }
\author{S.\ Viola Kusminskiy}\affiliation{Department of Physics, Boston University, 590 Commonwealth
  Ave., Boston, MA 02215}
\author{K.\ S.\ D.\ Beach}\affiliation{Institut f\"ur Theoretische Physik und Astrophysik, Universit\"at W\"urzburg, Am Hubland, D-97074 W\"urzburg, Germany}\affiliation{Deparment of Physics, University of Alberta, Edmonton, Alberta T6G 2D7, Canada}
\author{A.\ H.\ Castro Neto}\affiliation{Department of Physics, Boston University, 590 Commonwealth
  Ave., Boston, MA 02215}
\author{D.\ K.\ Campbell}\affiliation{Department of Physics, Boston University, 590 Commonwealth
  Ave., Boston, MA 02215}

\date{\today}

\begin{abstract}
We investigate the evolution of the heavy fermion ground state under application of
a strong external magnetic field. We present a richer version of the usual hybridization
mean field theory that allows for hybridization in both the singlet and triplet channels 
and incorporates a self-consistent Weiss field. We show that for a magnetic field strength
$B^\star$, a filling-dependent fraction of the zero-field hybridization gap,
the spin up quasiparticle band becomes fully polarized---an event marked by a sudden 
jump in the magnetic susceptibility.
The system exhibits a kind of quantum rigidity in which the susceptibility (and several
other physical observables) are insensitive to further increases in field strength.
This behavior ends abruptly with the collapse of the hybridization order parameter
in a first-order transition to the normal metallic state. We argue that the feature
at $B^\star$ corresponds to the ``metamagnetic transition'' in YbRh$_2$Si$_2$.
Our results are in good agreement with recent experimental measurements.
\end{abstract}


\maketitle

\section{Introduction}

The Kondo lattice model away from half filling describes what is called a
\emph{heavy metal}---a Fermi liquid state distinguished by its high magnetic 
susceptibility and large Sommerfeld coefficient. The properties of this state 
are a consequence of the hybridization between the localized impurity spins 
($f$ electrons) and the conduction electrons ($c$ electrons). The $f$
electrons are said to delocalize in the sense that the Luttinger volume
is now comprised of the sum of both species, $c$ and $f$.~\cite{Luttinger} The shallow 
dispersion of the resulting quasiparticles leads to a renormalization of the
effective mass to very large values.~\cite{Newns87} Materials that belong 
to this category are usually rare-earth intermetallic compounds, in which
$4f$ or $5f$ electrons act as magnetic impurities embedded in
a metallic host.~\cite{Stewart84}

It is known that a sufficiently strong magnetic field will destroy
the heavy fermion state. Measurements in applied field often
reveal a so-called metamagnetic transition (MMT), characterized by an 
abrupt change in the magnetic quantities. MMTs are ubiquitous among the 
different heavy fermion compounds, but due to the rich variety and 
complicated structure of their phase diagrams, the true nature of the 
MMT remains unknown. In particular, two competing scenarios are
under consideration. The localization scenario proposes that the
itinerant $f$ electrons suddenly localize at the MMT, and the heavy fermion
state disappears at that point. An alternative scenario
is given by a continuous evolution of the Fermi surface, in which the
localization of the $f$ electrons is not tied to the MMT.

Experimental results indicate the existence of MMTs in CeRu$_2$Si$_2$ 
and YbRh$_2$Si$_2$, which have been attributed to the localization of the $f$
electrons.~\cite{Tokiwa05,Aoki93} Recent experiments
on CeRu$_2$Si$_2$, however, suggest that the localization scenario
might not be appropriate.~\cite{Daou06} This was already pointed out in
Ref.~\onlinecite{Sakakibara95}, in which static magnetization measurements 
ruled out the possibility of a first order phase transition, at odds with the 
existent de~Haas--van~Alphen measurements.~\cite{Aoki93} On the basis of Hall
effect and magnetoresistance measurements, the authors of Ref.~\onlinecite{Daou06} 
proposed that the MMT anomalies result from a continuous evolution of the Fermi
surface in which the spin-split sheet corresponding to the heaviest electrons
shrinks to a point.

This paper elaborates on the continuous scenario.
We argue that the underlying phenomenology is the systematic descent of the spin up 
quasiparticle band with applied field. The MMT corresponds to the point at which 
the entire spin up band drops below the chemical potential, and only at much higher fields 
do the $f$ electrons localize. A common argument in favor of the localization scenario
is that a sharp drop in $\gamma$, the linear coefficient of the specific heat, is always
observed at the MMT. What renders this interpretation unconvincing is that the value of $\gamma$
above the MMT is too large,~\cite{Meulen91,Tokiwa05}
as is the residual magnetic susceptibility.~\cite{Tokiwa05}
A better picture is that as the field is ramped up, the mass enhancement 
factor of the spin up quasiparticles increases while that of the spin down decreases.~\cite{Beach05,Spalek90, Onari08}
At the MMT, the lower spin up band is completely filled, and the heaviest quasiparticles
drop out of all thermodynamic quantities: the system goes from having a 
spin up Fermi surface of very heavy particles and a spin down surface of moderately 
heavy particles to having only the latter.

We address the theoretical aspects of such a transition from a mean field point of view. 
Our approach goes well beyond the conventional $c$-$f$ hybridizidization 
scheme in that it treats \emph{all} the relevant competing interaction channels;
it includes both singlet and triplet pairing and allows for 
spontaneous ferromagnetism.
The particular compounds mentioned above are amenable
to this kind of analysis because they are largely paramagnetic: they do not 
exhibit superconductivity, and antiferromagnetic ordering is suppressed by 
very small values of an applied external magnetic field.
As to the overall reliability of the mean field approach, comparison with 
dynamical mean field simulations confirms that quantum fluctuations do not qualitatively 
change the behavior in the range of field strengths that are of interest for the MMT.~\cite{Beach08}

The starting point for our analysis is the periodic Anderson model~\cite{Anderson61} 
augmented by a Zeeman term.  In the limit of infinite on-site repulsion, a Schrieffer-Wolf
transformation maps the Hamiltonian to a model of conduction electrons
coupled to an array of localized magnetic impurities:
\begin{multline} \label{klm}
\hat{H} = 
    -t \sum_{\bracket{ij}}\bigl( c_i^{\dagger}c_j + c_j^{\dagger}c_i \bigr)\\
    +\sum_i\Bigl[ J\hat{\vec{s}}_i \cdot \hat{\vec{S}}_i
    -\mu_{\text{B}}\vec{B}\cdot \bigl(g_c \hat{\vec{s}}_i + g_f \hat{\vec{S}}_i \bigr)\Bigr].
\end{multline}
Here, $c_i^{\dagger} = (c_{i\uparrow}^{\dagger}\ c_{i\downarrow}^{\dagger})$ is the creation operator for
the conduction electrons and $\hat{\vec{s}}_i=\frac{1}{2} c^\dagger_i
\svec{\sigma}c_i$ is the operator that measures their local spin density.
$\hat{\vec{S}}_i$ is the impurity pseudo-spin at site $i$. It too has an underlying
fermion representation, $\hat{\vec{S}}_i = \frac{1}{2}f_i^{\dagger}\svec{\sigma}f_i$, but the
infinite on-site repulsion suppresses all charge fluctuations: i.e., $f_i^{\dagger}f_i=1$. 
The spinor components of $f$ refer to the appropriate Kramers doublet: 
the degeneracy of the physical $f$ electrons is effectively reduced to two by 
strong spin-orbit coupling and crystal field effects.~\cite{Lee86}
For Ce and Yb, these are the doublets of the 4f$^1$ and 4f$^{13}$ atomic configurations.

In Eq.~\eqref{klm}, Land\'{e} $g$-factors have been introduced
to allow for differential coupling of the $c$ and $f$ electrons to
the applied field $\mathbf{B}$. (In what follows, we set the Bohr magneton $\mu_{\text{B}} = 1$.)
The factor $g_f$ in particular is highly nontrivial and depends on the details of the 
$f$-electron environment. In general, such moments are partially quenched,~\cite{Zou86}
and simple estimates suggest that semi-realistic values are in the range
$1 \lesssim g_f \lesssim 1.5$ (see, e.g., Refs.~\onlinecite{Pietri05,Izumi07}),
while $g_c \approx 2$. 
In particular,  Ce$^{3+}$ ($J=5/2$, $L=3$, $S=1/2$) has $g_f=7/6$
and  Yb$^{3+}$ $J=7/2$ ($J=5/2$, $L=3$, $S=1/2$) has $g_f=8/7$.
The value of $g_f$ may also have some field dependence,~\cite{Pietri05} but this effect
is weak enough to ignore.

Some authors have worked in the limit $g_c = 0$ so that the
field couples only to the impurity spin and not at all to the conduction
electrons,~\cite{Ohkawa89,Hong92,Saso96,Ono96,Ono98,Satoh01}
but we do not believe that this is the correct starting point. A more common
assumption is to set the two $g$-factors equal.~\cite{Meyer01,Milat04,Beach04,Ohashi04}
This choice, however, is a somewhat artificial limit~\cite{Nakano91} and, at the mean field
level, leads to a nongeneric ($g_c = g_f$ is a special tuning) magnetization plateau 
of width equal to the zero-field Kondo energy.~\cite{Beach05,Beach08}

We do not attempt to model the anisotropy of $g$ (it is in principle a
tensor~\cite{Sichelschmidt03}) by introducing explicit crystal field
terms into the Hamiltonian.~\cite{Konno91} 
We disregard the fact that metamagnetism appears when the field is applied 
parallel to an easy axis (the tetragonal c-axis~\cite{Puech88}). Another simplification in our model is that we do not consider the Ruderman-Kittel-Kasuya-Yosida (RKKY) interaction, and therefore it is valid for paramagnetic ground states in which long range magnetic correlations are absent. Neglecting the RKKY interaction in our case is \emph{a priori} justified since we are interested in the high external magnetic field regime.

In the next section we present the mean field construction for this Hamiltonian 
and derive the self-consistent equations that determine the mean field parameters.
We show that the mean field must include separate hybridization parameters in the
spin up and spin down channels and a Weiss molecular field in order for the model to capture all the observed qualitative features of the system.
Results from solving for the parameters are presented in
Sec.~\ref{sec:results}. In particular, we show the evolution of the Kondo gap, mass enhancement factor,
magnetization and susceptibility of the system for increasing external
magnetic field, and we sketch the phase diagram predicted by the model. 
We discuss our results in Sec.~\ref{sec:discuss} and provide a summary of our principal results in
Sec.~\ref{sec:summ}. Some details of the calculations are relegated to the appendices.

\section{Mean Field Approach}\label{sec:MF}

A hybridization mean field treatment of the KLM Hamiltonian was presented in 
Ref.~\onlinecite{Beach05} based on a decomposition in terms of the operators 
$\hat{\chi}^{\mu}=\frac{1}{\sqrt{2}}f^{\dagger}\sigma^{\mu}c$. The index
$\mu$ ranges over $0,1,2,3$; in this notation, $\sigma^0$ is the $2\times 2$ identity matrix 
and $\sigma^1,\sigma^2,\sigma^3$ are the Pauli matrices.
These operators are complete ($\sum_{\mu} \chi^{\mu\dagger}\chi^{\mu} = 1$)
in the $\tfrac{1}{2} \otimes \tfrac{1}{2} = 0\oplus 1$ spin sectors and are
introduced so that the exchange interaction can be explicitly
broken up into singlet and triplet components:
\begin{equation} \label{EQ:ExchangeInteraction}
\frac{1}{4}c^{\dagger}\svec{\sigma}c \cdot f^{\dagger}\svec{\sigma}f= -\frac{3}{4} \hat{\chi}^{0\dagger}\hat{\chi}^0
+ \frac{1}{4} \hat{\svec{\chi}}^\dagger \cdot \hat{\svec{\chi}}.
\end{equation}
Here, the three triplet components are represented using vector notation, $\hat{\chi}^{\mu} = (\hat{\chi}^0,\hat{\svec{\chi}})$.

In the usual way, the right-hand side of Eq.~\eqref{EQ:ExchangeInteraction}
can be approximated by
\begin{equation} \label{pairch}
\sum_{\mu}\biggl(\frac{1}{4}-\delta^{\mu 0}\biggr)\biggl( \chi^{\mu\dagger}\bracket{\chi^{\mu}}
- \bracket{\chi^{\mu}}^*\chi^{\mu} - \bigl\lvert \bracket{\chi^{\mu}} \bigr\rvert^2 \biggr).
\end{equation}
In zero field, it is customary to assume that only the singlet amplitude condenses
($\bracket{\hat{\chi}^0}\neq 0$ 
and $\bracket{\hat{\svec{\chi}}}=0$). This gives rise to the conventional heavy fermion state. 
At nonzero field, however, it is appropriate to consider the possibility of a nonzero
\emph{triplet} hybridization. For convenience, we introduce a spin-dependent hybridization energy,
\begin{alignat}{2}\nonumber
        V^0 &= \frac{3J}{4\sqrt{2}} \bracket{\hat{\chi}^0} 
&\qquad  V^3 &= \frac{J}{4\sqrt{2}} \bracket{\hat{\chi}^3}\\
        V_+ &= V^0 - V^3
&       V_- &= V^0 + V^3.
\end{alignat}
These definitions are consistent with an applied magnetic field $\vec{B} = (0,0,B)$
directed along the axis of spin quantization. Then Eq.~\eqref{pairch},
the decomposition in the pairing channel, can be expressed as
\begin{equation} \label{EQ:PairingDecomposition}
\begin{split}  
J\hat{\vec{s}} \cdot \hat{\vec{S}} \,\xrightarrow{\text{pair.}}\, & -V^{0*} f^\dagger c
- V^0 c^\dagger f + \frac{8|V^0|^2}{3J}
\\ & +V^{3*} f^\dagger \sigma^3 c+ V^3 c^\dagger \sigma^3 f - \frac{8|V^3|^2}{J}.
\end{split}
\end{equation}
The triplet hybridization, which to our knowledge has never been included in any mean
field treatment, plays an important role in the vicinity of the metamagnetic transition.
Admitting the possibility of $V_{\uparrow} \neq V_{\downarrow}$
makes our theory compatible with the quasiparticle interpretation of
the Gutzwiller approach.~\cite{Reynolds92}

We also decompose the Kondo interaction in the magnetic channel. That is,
\begin{equation} \label{EQ:MagneticDecomposition}
\hat{\vec{s}} \cdot \hat{\vec{S}} \,\xrightarrow{\text{mag.}}\,
\hat{\vec{s}} \cdot \vec{m}_f + \vec{m}_c \cdot \hat{\vec{S}} - \vec{m}_c\cdot\vec{m}_f,
\end{equation}
where $\vec{m}_c = \bracket{\hat{\vec{s}}}$ and $\vec{m}_f = \bracket{\hat{\vec{S}}}$.
This sets up a Weiss molecular field whereby every $c$ electron feels
the counter-polarizing effect of its local $f$ partner, and vice versa. 
Such a contribution is necessary to reproduce the observed diamagnetism of the $c$ electrons, 
which initially cant away from the applied field.~\cite{Beach04,Ohashi04}

Summing the contributions of Eqs.~\eqref{EQ:PairingDecomposition} and \eqref{EQ:MagneticDecomposition}
yields a complete mean field Hamiltonian,
\begin{equation}
\begin{split}
\hat{H}_{\text{\tiny MF}}= 
   & -t\sum_{\bracket{ij}}\bigl(c_i^{\dagger}c_j+c_j^{\dagger}c_i\bigr)
     -\sum_{is} \bigl(V_sc_{is}^{\dagger}f_{is}+V_s^*f_{is}^{\dagger}c_{is}\bigr)\\
   & -\sum_{is}\sum_{a=c,f} a^\dagger_{is}a_{is} \Bigl[ \mu_a + \frac{s}{2} \bigl(g_a B-m_{\bar{a}} J\bigr) \Bigr] 
     + N\mathcal{F}_0.
\end{split}
\end{equation}
The indexes $i$,$j$ run over the lattice sites, $N$ is the total number
of sites, and $s=\pm 1$ corresponds to spin up and down. The bar indicates
$\bar{a}=f$ ($c$) if $a=c$ ($f$). The overall constant is given by
\begin{equation} \label{F0}
\mathcal{F}_0 = \frac{4}{3J}\Bigl( 4V_+ V_--V_+^2 - V_-^2 \Bigr) - J m_cm_f.
\end{equation}
Note that we have introduced chemical potentials $\mu_c$ and $\mu_f$ in order to control 
the occupation of the $c$ and $f$ electrons; the constraints 
$\bracket{f_i^{\dagger}f_i}=n_f\equiv 1$ and $\bracket{c_i^{\dagger}c_i}=n_c$ 
are imposed on average. This mean field decomposition is justified by a variational argument, detailed in 
Appendix~\ref{sec:appA}, which uniquely determines the numerical prefactors appearing
in Eq.~\eqref{F0}.

The mean field Hamiltonian can then be written in Fourier space as
\begin{equation}\label{fouriermfh}
H_{\text{\tiny MF}} =
   \sum_{\vec{k}s}\begin{pmatrix} c_{\vec{k}s}^{\dagger} & f_{\vec{k}s}^{\dagger}\end{pmatrix}
   \mathrm{M}_{\vec{k} s}\begin{pmatrix} c_{\vec{k}s}\\ f_{\vec{k}s}\end{pmatrix} + N\mathcal{F}_0
\end{equation}
with coefficient matrix
\begin{equation}
\mathrm{M}_{\vec{k}s}= \begin{pmatrix}
\ek -\mu_c-\frac{s}{2}g_cB_c & -V_s\\
-V_s & -\mu_f-\frac{s}{2}g_f B_f 
\end{pmatrix}.
\end{equation}
Here, $\ek$ is the dispersion relation of the bare $c$ electrons and
$B_f= B-Jm_c/g_f$,
$B_c= B-Jm_f/g_c$ are self-consistent Weiss fields.
Note that the effective field felt by an $a$-electron ($a = c,f$)
differs considerably from the applied field when $J$ is large and $g_a$ is small.

The eigenvalues of $\mathrm{M}_{\vec{k}s}$ are $E_{\vec{k}s}^n=I_{\vec{
    k}s}^n-\mu_f-\frac{s}{2}g_f B_f$,
where $n=\pm 1$ is a quasiparticle band index and we have defined 
\begin{equation}\label{Iks}
I_{\vec{k}s}^n=\frac{1}{2}\Bigl[ \ek - b_s
  +n\sqrt{(\ek -b_s)^2+4V_s^2} \Bigr]
\end{equation}
with $b_s = b+(s/2)(g_cB_c - g_fB_f)$ and
$b=\mu_c-\mu_f$. $b$ is the chemical energy
for transmuting $c$ electrons into $f$ electrons;
it becomes increasingly important away from half filling.

In terms of the free energy $\mathcal{F}= \mathcal{F}_0 -\frac{1}{N \beta}\sum_{\vec{k}ns}
\ln\bigl(1+e^{ -\beta E_{\vec{k}s}^n}\bigr)$, the mean field values are
determined by solving the following system of equations:
\begin{alignat}{3}\label{constraints}\nonumber
-\frac{\partial \mathcal{F}}{\partial \mu_c}&=n_c &\qquad
-\frac{1}{g_c}\frac{\partial \mathcal{F}}{\partial B_c}&=m_c &\qquad
-\frac{\partial \mathcal{F}}{\partial V_-}&=0 \\
-\frac{\partial \mathcal{F}}{\partial \mu_f}&=n_f &\qquad
-\frac{1}{g_f}\frac{\partial \mathcal{F}}{\partial B_f}&=m_f &
-\frac{\partial \mathcal{F}}{\partial V_+}&=0.
\end{alignat}
It is understood that
\begin{alignat}{2}\nonumber
n_{c\uparrow}+n_{c\downarrow}&=n_c & \qquad 
n_{f\uparrow}+n_{f\downarrow}&=n_f \\
n_{c\uparrow}-n_{c\downarrow}&=2m_c & \qquad 
n_{f\uparrow}-n_{f\downarrow}&=2m_f.
\end{alignat}

Equations~\eqref{constraints} can be translated to the continuum by defining
$D_s(\omega)= \sum_{{\vec{k}}n} \delta(\omega-I_{{\vec{k}}s}^n)$, 
a density of states (DOS) shifted with respect to the true energy
zero by $\mu_{fs}\equiv\mu_f+\frac{s}{2}g_fB_f$. Here, we can now follow 
the method described in Ref.~\onlinecite{Beach05}, extending it to take the spin dependence into account.
The result is as follows. If we assume that the bare conduction-electron DOS
has the form
\begin{equation}
\sum_{{\vec{k}}}\delta(\omega-\ek)=\phi(\omega)\theta( W^2-4\omega^2),
\end{equation}
where $\phi(\omega)$ describes the spectral line shape and the Heaviside
function $\theta$ demarcates the band edges at $-W/2$ and $W/2$,
then the renormalized $c$-electron DOS is
\begin{equation}
D_s^c(\omega) = \phi\Bigl(\omega-\frac{V_s^2}{\omega}+b_s\Bigr)\sum_{l=1}^4\theta(\omega-\omega_{ls}),
\end{equation}
where the edges of the quasiparticle
dispersion bands are now given by the expressions
\begin{equation}\label{ws}
\begin{split}
\omega_{1s}=&-\frac{1}{4}\sqrt{(W+2b_s)^2+(4V_s)^2}-\frac{1}{4}(W+2b_s)\\
\omega_{2s}=&-\frac{1}{4}\sqrt{(W-2b_s)^2+(4V_s)^2}+\frac{1}{4}(W-2b_s)\\
\omega_{3s}=&+\frac{1}{4}\sqrt{(W+2b_s)^2+(4V_s)^2}-\frac{1}{4}(W+2b_s)\\
\omega_{4s}=&+\frac{1}{4}\sqrt{(W-2b_s)^2+(4V_s)^2}+\frac{1}{4}(W-2b_s).
\end{split}
\end{equation}
The total quasiparticle DOS is given by
\begin{equation}
D_s=D_s^c(\omega)+D_s^f(\omega) \quad \mbox{with}
\quad D_s^f(\omega)=\frac{V_s^2}{\omega^2}D_s^c(\omega).
\end{equation}

The Kondo gap depends on spin and is given by
$\Delta_{Ks}=\omega_{3s}-\omega_{2s}$. It is straightforward to see that
$\Delta_{Ks}$ collapses to zero when $V_s=0$, and therefore it can be thought of
as an alternative order parameter for the heavy fermion state.

One finds that Eqs.~\eqref{constraints} are equivalent to (for each of $a=c,f$)
\begin{equation}\label{constcont}
\begin{split}
\sum_s\int\!d\omega\,f(\omega-\mu_{fs})D_s^{a}(\omega)&=n_{a}\\
\sum_s s\int\!d\omega\,f(\omega-\mu_{fs})D_s^{a}(\omega)&=n_{a \uparrow}-n_{a
  \downarrow}\\
\frac{3J}{4} \int\,d\omega\,f(\omega-\mu_{fs})\frac{D_s^c(\omega)}{\omega}&=1-\frac{2V_{-s}}{V_s},
\end{split}
\end{equation}
where $f(\omega)$ denotes the Fermi function. In the special case of a 
flat $c$-electron spectrum $\phi = 1/W$ and zero temperature $f(\omega-\mu_{fs})\to \theta(\mu_{fs}-\omega)$,
these integral equations have a closed form~\cite{Lacroix79,Beach05} and thus can be solved efficiently everywhere
in the $J,B,n_c$ parameter space.

\begin{figure*}
\includegraphics{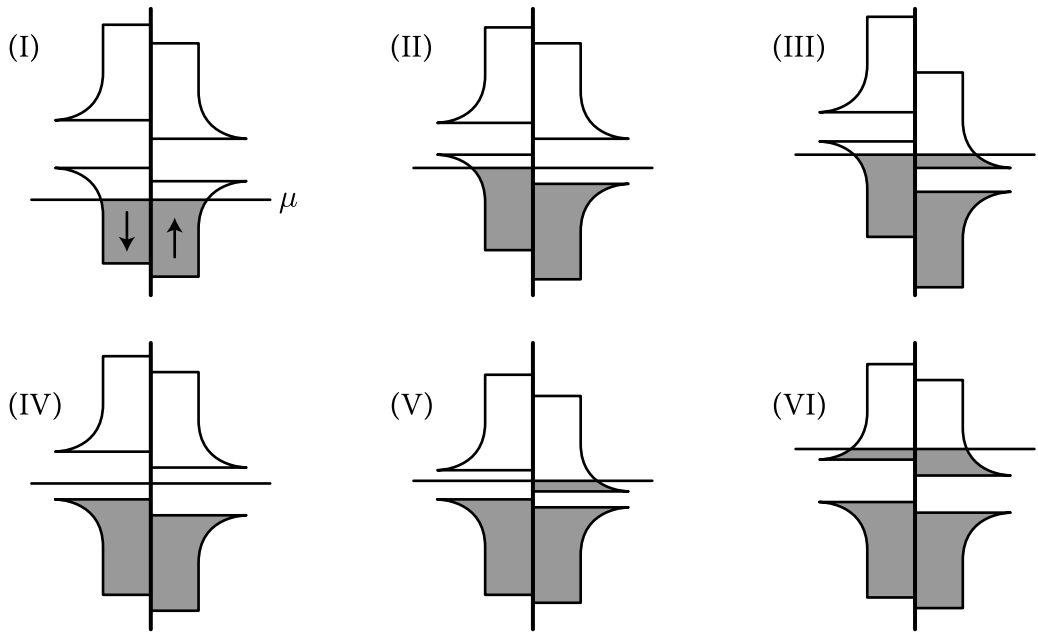}
\caption{\label{muf} The six possible placements of the bands with respect to the
chemical potential. Cases (I) and (II) occur for $0 < n_c < 1$ and cases (V)
and (VI) for $1 < n_c < 2$. Case (IV) occurs for the $n_c=1$ Kondo insulator.
We find that the controversial case (III) never occurs: it either has no solution or a 
solution that is energetically unfavorable.
}
\end{figure*}

For $B>0$, the important issue is where $\mu_{f+}$ sits with respect to the band edges.
There are six possibilities:
\begin{alignat*}{4}
&\text{(I)} &\quad \omega_{1-} < \mu_{f-} < \omega_{2-} 
&\quad \text{(II)} &\quad \omega_{1-} < \mu_{f-} < \omega_{2-}\\
& & \omega_{1+} < \mu_{f+} < \omega_{2+} 
& & \omega_{2+} < \mu_{f+} < \omega_{3+}\\
\\
&\text{(III)} &\quad \omega_{1-} <  \mu_{f-} < \omega_{2-} 
&\quad \text{(IV)} &\quad  \omega_{2-} < \mu_{f-} < \omega_{3-}\\
& & \omega_{3+} < \mu_{f+} < \omega_{4+} 
& & \omega_{2+} <  \mu_{f+} < \omega_{3+}\\
\\
&\text{(V)} &\quad \omega_{2-} < \mu_{f-} < \omega_{3-}
&\quad\text{(VI)} &\quad \omega_{3-} < \mu_{f-} < \omega_{4-}\\
& & \omega_{3+} < \mu_{f+} < \omega_{4+}
& & \omega_{3+} <  \mu_{f+} < \omega_{4+}
\end{alignat*}

In zero field, if the system is not spontaneously magnetic, we have a situation
where $\mu_{f+} = \mu_{f-}$ sits either inside the lower hybridized bands ($0 < n_c < 1$)
or inside the upper hybridized bands ($1 < n_c < 2$). We will restrict our attention to
the case of less than half filling. (If the bare band structure is symmtric then
the two cases are equivalent.) When the external magnetic field is turned on,
$\mu_{f-}$ will descend and $\mu_{f+}$ will ascend. For small fields
($B \ll \Delta_{Ks}$), the chemical potential of the spin up quasiparticles will
still sit below the upper edge of the lower band ($\mu_{fs} < w_{2s}$). At larger
fields, we have to account for the possibility of $\mu_{f+}$ moving into the hybridization
gap or into the upper quasiparticle band (positions II and III in Fig.~\ref{muf}).
All of these possibilities have to be solved for, along with the $V_s=0$ case, with
the true ground state determined by energy considerations.

Some authors have ascribed the metamagnetic transition to a regime II~$\to$~III crossover~\cite{Nakano91,Hong92,Hong92b} or even to a direct regime I~$\to$~III transition.~\cite{Edwards97} Our model predicts, however, that the origins of metamagnetism lie in the smooth crossover from regime I to II. What we find is that at some
field $B^\star$, $\mu_{f+}$ moves smoothly into the gap so that $\mu_{f+}$ is replaced
by $\omega_{2+}$ in the upper limits of the integrals in Eqs.~\eqref{constcont}. Hence the only explicit magnetic field dependence in the system of equations 
comes through $\mu_{f-} = m_f-\tfrac{1}{2}g_fB_f$. $\mu_{f-}$ itself becomes
field independent ($\mu_f = \text{const} + \tfrac{1}{2}g_fB_f$) and all the
quantities that depend on it become locked at the values obtained at $B^\star$.
In the special case of $g_c = g_f$, $B = B^\star$ marks the sudden collapse of the
magnetic susceptibility to zero and the beginning of a large magnetic plateau.
For general $g_c \neq g_f$, we instead see the susceptibility jump to a lower
but nonzero value; the magnetization changes slope and becomes perfectly linear
for $B > B^\star$.

The crossover to the ``locked'' regime II is determined by the balance of the
free energies evaluated at the two sets of solutions, if existing. The free
energy density for the heavy fermion state can be calculated from
its expression in the continuum:
\begin{equation}\label{freeEcont1}
\mathcal{F}= \mathcal{F}_0 -\frac{1}{\beta}\sum_s\int d\omega
D_s(\omega) \ln{\Bigl[1+e^{-\beta(\omega-\mu_{fs})}\Bigr]}.
\end{equation}
Using 
$\mathcal{F}\!=\!\partial (\beta \mathcal{F})/\partial \beta$ 
and the self-consistent
equations for $V_s$, we can obtain the functional expression for
$\mathcal{F}(\mu_c,\mu_f,B)$ within the different regimes. For example, in regime I:
\begin{equation}\label{freeEcont2}
\begin{split}
{\mathcal F_I}=& 
-J m_cm_f-g_fB_f(m_c+m_f)-\mu_f(n_c+n_f)\\ & +\frac{1}{2W}\sum_s\bigl( \mu_{fs}^2 - \omega_{1s}^2\bigr).
\end{split}
\end{equation}
From the definition of the free energy 
$\mathcal{F}(\mu_c,\mu_f,B)= \mathcal{U}(n_c,n_f,m)-Bm-\mu_cn_c-\mu_fn_f$, we can also compute the Gibbs free energy $\mathcal{G}(n_c,n_f,B)=\mathcal{U}-Bm$. Here we have written the total magnetization $m=g_cm_c+g_fm_f$, and it is understood that we are working at zero temperature. It
is convenient to introduce a quantity $x$, defined by $n_c=1-x$, that measures the deviation from half filling.
Since $n_f=1$ is fixed, we can write $\mathcal{G}(x,B)=\mathcal{F}(\mu_c,\mu_f,B)+\mu_c(1-x)+\mu_f$. For regime I, we find
\begin{equation}\label{GibbsEcont}
\begin{split}
{\mathcal G_I}=& b(1-x)- B g_f(m_c+m_f) + Jm_c^2\\ &+\frac{1}{2W}\sum_s\bigl( \mu_{fs}^2 - \omega_{1s}^2\bigr).
\end{split}
\end{equation}
The free energy for the locked
regime II is obtained by substituting $\mu_{f+} \to \omega_{2+}$ in
Eqs.~\eqref{freeEcont2} and \eqref{GibbsEcont}.

In the limit $V_+ = V_-$ and $B = B_c = B_f$, which corresponds to the simpler mean field theory 
described in Ref.~\onlinecite{Beach05}, this reduces to
\begin{equation}
{\mathcal G}= \frac{1}{W} \Bigl( \mu_f^2 + \frac{1}{4}B^2
- \omega_{1}^2 \Bigr) - B(m_c+m_f) + b(1-x).
\end{equation}
The heavy fermion state will
collapse when the free energy $\mathcal{G}$ of the heavy fermion (in phase I, II, or III)
is greater than either the free energy ${\mathcal G_{PM}}$ of a normal, paramagnetic metallic state 
or the free energy ${\mathcal G_{IF}}$ of a conventional itinerant ferromagnet. 
This transition occurs at a critical field $B^{\text{c}}$ such that either ${\mathcal G}={\mathcal G_{PM}}$ or ${\mathcal G}={\mathcal G_{IF}}$, whichever is smaller. The expressions for these energies are given in 
Appendix~\ref{sec:appB}, where we also present the criterion for determining which end-state wins out.

\section{Results}\label{sec:results}
\begin{figure}
\includegraphics[scale=.33,angle=270,keepaspectratio=true]{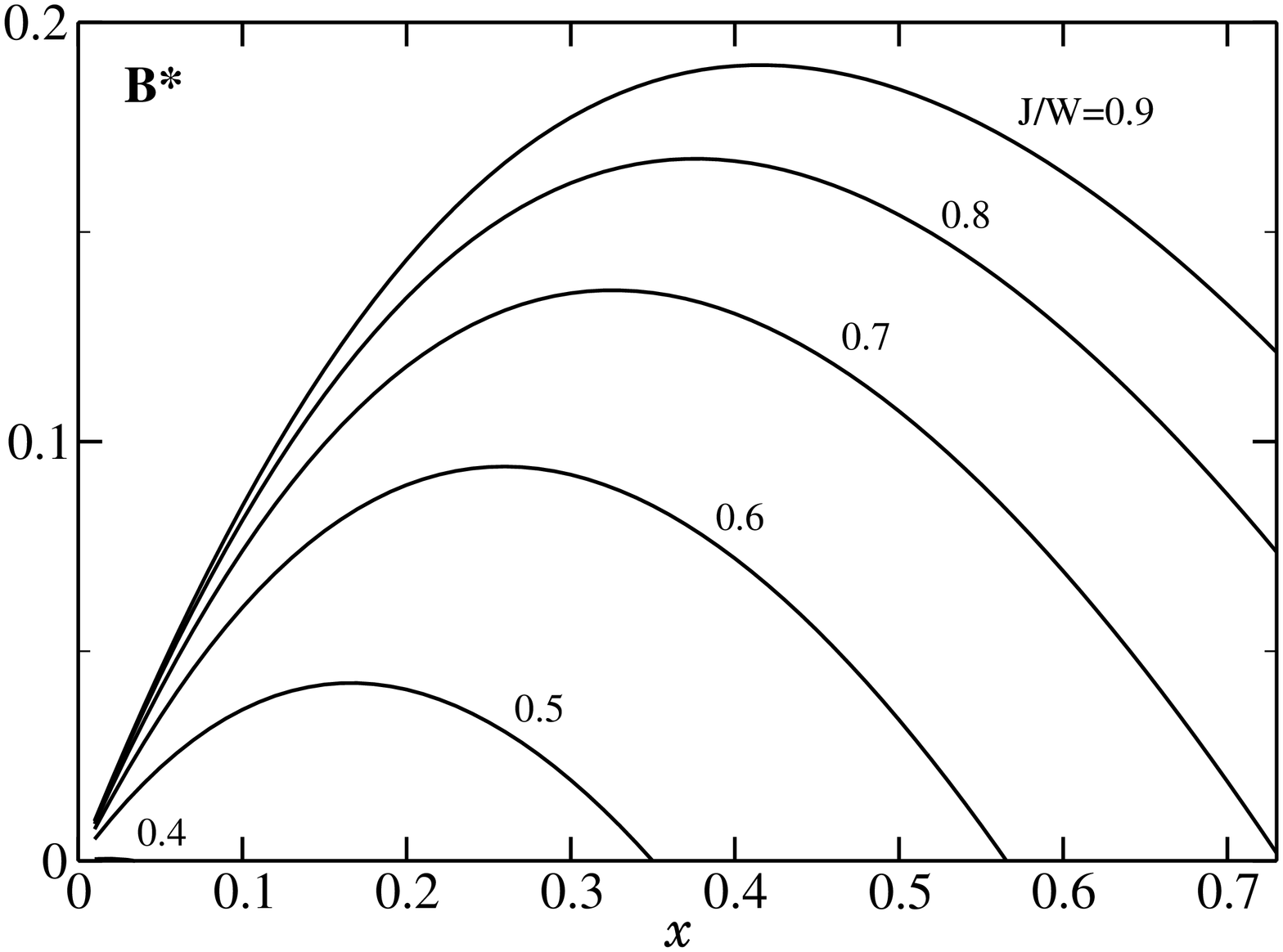}
\includegraphics[scale=.33,angle=270,keepaspectratio=true]{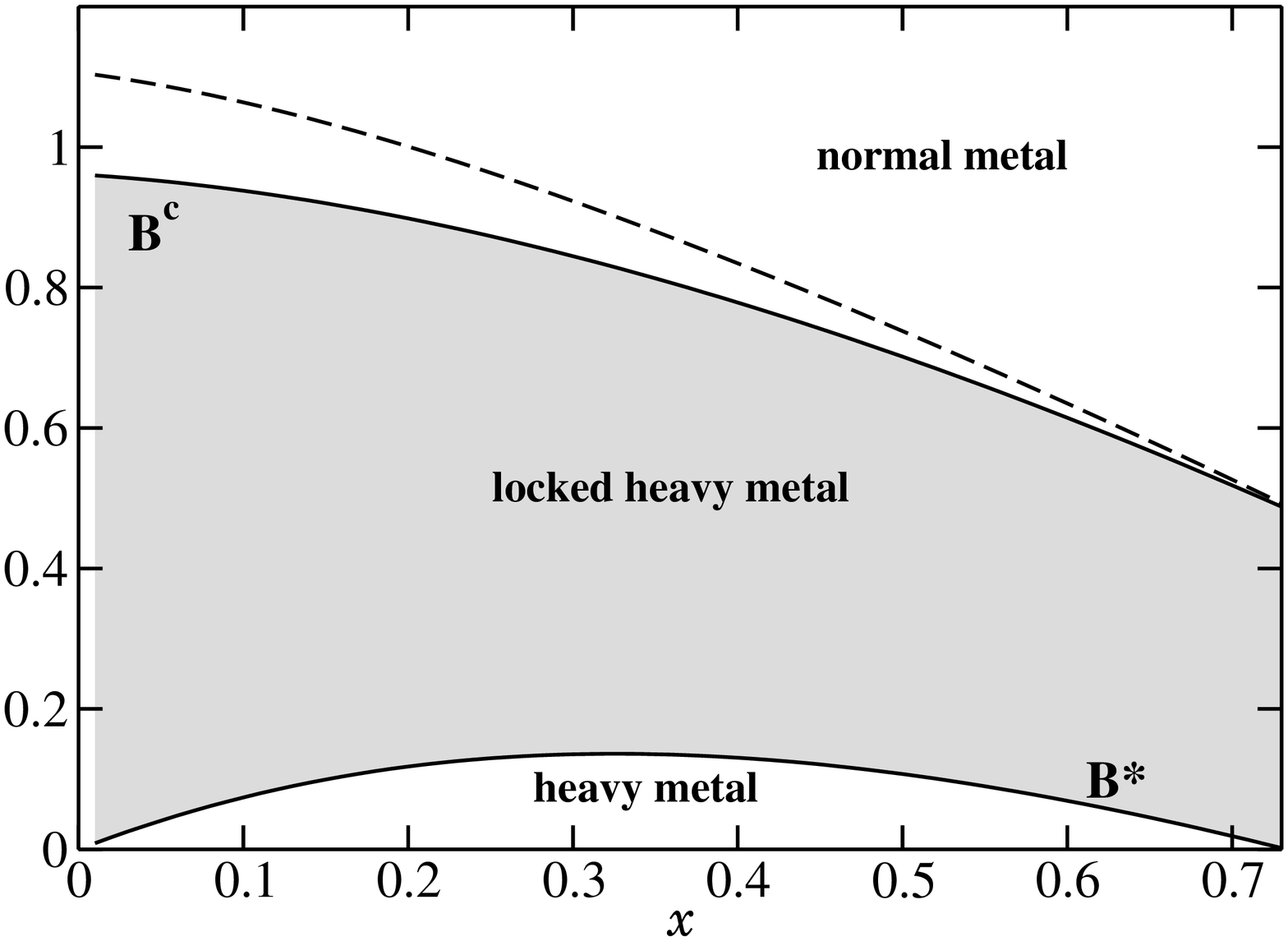}
\caption{\label{fig:phases}
(Top) The crossover field $B^\star$ at which $\mu_{f+}$ moves into the hybridization
gap is plotted in units of the zero-field gap as a function of the filling parameter $x$
for various values of the Kondo coupling. We have chosen $g_c = 2$, $g_f=8/7$.
(Bottom) $J/W=0.7$. $B^\star$ is the MMT and $B^{\text{c}}$ is the critical field  of the
first-order transition back to the normal state. The dashed line marks the point
at which $\mu_{f+}$ would have entered the upper quasiparticle band if
it had not been preempted by the destruction of the heavy state.}

\end{figure}

The system of equations~\eqref{constcont} was solved numerically by turning on adiabatically the external magnetic field B, for
different values of the Kondo coupling $J$ and filling factor $x$. A typical phase diagram obtained with our model is shown in the bottom plot of Fig.~\ref{fig:phases} for a particular value of the exchange coupling $J$ and as a function of the filling factor $x$. We see that there is always a magnetic field $B^*<B^{\text{c}}$ for which the system enters the locked state, regime II. As we discussed in the previous section, $B^*$ corresponds to the magnetic field for which $\mu_{f+}=\omega_2$. Between $B^*$ and $B^{\text{c}}$, $\mu_{f+}$ is inside the gap and, as we discussed earlier, 
the set~\eqref{constcont} has a field-independent
solution. $B^{\text{c}}$ is the critical magnetic field at which the heavy fermion state is destroyed. From the diagram we also see that the heavy fermion state always collapses before regime III can be reached. 

The top plot of Fig.~\ref{fig:phases} shows the behavior of $B^*$ as a function of the filling factor $x$ for different values of $J$. As the coupling $J$ decreases, the heavy metal phase is constrained to smaller values of $x$.  It is interesting to note that while the field $B^{\text{c}}$ is a monotonic decreasing function of $x$,  $B^*$ has a dome shape  and it is maximum for $x$ at approximately the midpoint of the range for the regime I phase determined for each $J$. The behavior of $B^{\text{c}}$, which is not shown in this plot, is similar to the behavior depicted in the phase diagram for $J/W=0.7$. It is important to remark that $B^{\text{c}}$ is always at least one order of magnitude larger than $B^*$.
To the right of the dome enclosing the regime I heavy metal, the system is already in the locked phase
at $B=0$, implying a spontaneous polarization. This can be related to the existence of strong ferromagnetic
correlations at low $n_c$.~\cite{Lacroix79,Yamamoto90} It is known that a ferromagnetic state exists in the 
limit of a single conduction electron, $n_c \to 0^+$ ($x \rightarrow 1^-$).~\cite{Sigrist91}

\begin{figure}
 \centering
 \includegraphics[scale=.33,angle=270,bb=8 54 665 710]{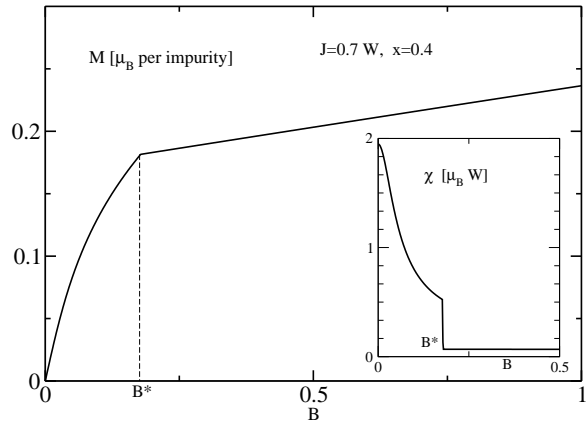}
 \caption{(Main) Total magnetization of the system with $g_f=8/7$, $g_c=2$ for $J/W=0.7$ and filling factor $x=0.4$, as a function of the magnetic field in units of the zero field gap. $B^*$ is the MMT. The critical field $B^{\text{c}}$ is out of the plot range. (Inset) Total susceptibility.}
 \label{fig:MandChi}
\end{figure}

By fixing both $J$ and $x$, we can plot the total magnetization of the system as shown in Fig.~\ref{fig:MandChi} for $J/W=0.7$ and $x=0.4$. The total magnetization is defined as $m=g_cm_c+g_fm_f$, $m_c$ giving a diamagnetic contribution. We see that there is a strong, nonlinear increase of the magnetization up to $B=B^*$. At this value of the external field, there is an abrupt change signaled by a kink, after which the magnetization is purely
linear with the field. At larger values of the field (not shown in the figure), another change would be expected at $B=B^{\text{c}}$ due to the collapse of the heavy fermion state. The inset of Fig.~\ref{fig:MandChi} shows the total susceptibility, which decreases in a nonlinear fashion up to $B^*$ and then jumps down abruptly to a constant. These plots are in remarkable qualitative agreement with the magnetic behavior found recently for YbRh$_2$Si$_2$, as can be seen by comparing with Ref. \onlinecite{Tokiwa05}.

\begin{figure}
 \begin{center}
 \includegraphics[scale=.33,angle=270,keepaspectratio=true,bb=8 54 665 710]{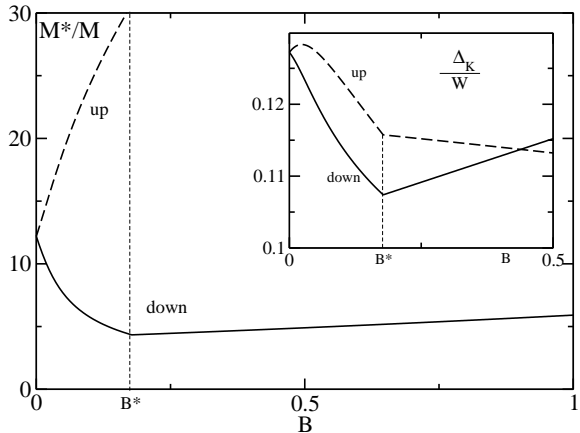}
\caption{\label{fig:mass}(Main) Mass enhancement factor for spin up (dashed) and spin down (solid) bands with $g_f=8/7$, $g_c=2$ for $J/W=0.7$ and filling factor $x=0.4$, as a function of the magnetic field in units of the zero field gap. $B^*$ is the MMT. The critical field $B^{\text{c}}$ is out of the plot range. (Inset) Kondo gap for spin up (dashed) and spin down (solid) bands.}
\end{center}
\end{figure}

The mass enhancement of the quasiparticles can be computed via the expression $(M^*/M)=\partial I_{ks}^n/\partial \epsilon_{ks}$ evaluated at the Fermi surface, where $M^*$ is the effective mass of the quasiparticles while $M$ is the mass given by the noninteracting bands. The result, again for $J/W=0.7$ and $x=0.4$, is shown in Fig.~\ref{fig:mass}. As predicted, the effective mass of the spin up (down) band increases (decreases) with increasing field. For $B>B^*$, however, the spin up band becomes infinitely massive while the spin down effective mass becomes roughly a constant. The inset of Fig.~\ref{fig:mass} shows the splitting of the Kondo gap due to the magnetic field.

The locked state can only be avoided if the heavy fermion state collapses before the locked state is reached.
This can be engineered for a small range of the parameters $J$ and $x$, if we assume $g_f<1$. This assumption,
however, is not justified by our simple estimates of $g_f$. Also, the very constrained range of parameters 
for which this behavior is obtained within our model would imply a high degree of fine tuning.

\section{Discussion}\label{sec:discuss}

Our model seems to capture qualitatively the physics behind the MMT of YbRh$_2$Si$_2$, indicating that the MMT is due to the crossover in which the spin up band ceases to participate in the heavy fermion state. Meanwhile, the spin down band contributes with a moderately large effective mass. The corresponding magnetic susceptibility is a moderately large constant value, as observed in YbRh$_2$Si$_2$.~\cite{Tokiwa05} The collapse of the heavy fermion phase is reserved for much higher fields, at least one order of magnitude larger than the MMT field. This is in agreement with observations for CeRu$_2$Si$_2$, which locate the MMT at approximately 10$\,$T, while the complete suppression of the heavy fermion state is not realized before fields on the order of 100$\,$T.~\cite{Meulen91}  We emphasize that the magnetic field strength required to polarize the quasiparticles 
of the renormalized heavy fermion bands is not on the order of the bandwidth, as is the case 
in the non-interacting ($J=0$) system. This is so due to the formation of hole-pockets at the top of 
the lower band, with characteristic energies of the order of the Kondo temperature, which are 
inherent to the hybridization process.~\cite{Denlinger01}

There is a qualitative difference between the MMT transition of YbRh$_2$Si$_2$ and the one of CeRu$_2$Si$_2$. While for the Yb compound the magnetization presents a rather smooth kink and the magnetic susceptibility jumps down, for the Ce compound there is a pronounced peak in the susceptibility at the MMT. According to our model, there is no such peak at $B^*$, altough a peak would be observed in the susceptibility at $B^{\text{c}}$.  However this would imply that the heavy fermion state is destroyed, and the remnant heavy fermion behavior would then be unexplained. This apparent contradiction was previously pointed out in Ref.~\onlinecite{Daou06}. A very similar behavior to that of CeRu$_2$Si$_2$ is observed for another heavy fermion compound, UPt$_3$.~\cite{Meulen90} Moreover, a spin-split surface was observed in UPt$_3$, with one of the components surviving with high mass after the MMT.~\cite{Julian92} The peak in the susceptibility in these compounds is usually attributed to enhanced antiferromagnetic fluctuations~\cite{Meulen90,Meulen91,Daou06,Tokiwa05}  and therefore it would be reasonable to expect that RKKY interactions, which are neglected in our model, play an important role at the MMT of these materials.

In order to compare quantitatively with the experimental data, we need to fix the free parameters: the Kondo coupling $J$, the filling factor $x$ and the bandwidth $W$. This is not trivial since these are noninteracting parameters, and therefore their true value is unknown. This is also true for the gyromagnetic factors $g_c$ and $g_f$, which we will take as in the previous section, $g_f=8/7$, $g_c=2$. We begin by fixing $B^*$, by imposing that it coincides with the MMT. Therefore we
identify $B^* \approx 10$ T, as a reasonable order of magnitude
suggested by the available experimental data.~\cite{Tokiwa05,Aoki93} Chosing a value for the noninteracting
bandwith $W$ restricts the range of $x$ and $J$. For $W \approx 1$~eV $\approx 10^4$~K $\approx 10^4$~T, 
we find that $B^*/W \approx
10^{-3}$. To fix the values of $J$ and $x$ we are guided the phase diagrams depicted in Fig.~\ref{fig:phases}. For attaining such small values of $B^*/W$, we see that $x$ has to be either small or close in value to $J/W$. Since the fraction of conduction electrons controls the magnetization, bigger $x$ will result in higher values of magnetization. Choosing $x=0.69$ for $J/W=0.7$ gives a total magnetization of $m \approx 0.5 \mu_B$ per impurity at the MMT, which is of the same order of magnitude as the observed for YbRh$_2$Si$_2$. However, the residual susceptibility after the MMT is two orders of magnitude less than the experimentally observed, and the Kondo gap is approximately 100K, which is four times bigger than the Kondo temperature for the material.~\cite{Gegenwart02}

However for $x=0.4$ and $J/W=0.7$, we get a Kondo gap of $\approx$ 10K, a magnetization of $m \approx 0.2 \mu_B$ per impurity at the MMT,  and a susceptibility of $\approx 0.01 \mu_B$/T per impurity, which are all of the order of magnitude of the experimental data. The drawback is assuming a noninteracting bandwidth of $W \approx 10^{-2}$ eV, which seems unphysical. Nonetheless, this kind of scale problem seems to be ubiquitous among the various Kondo lattice models; see, for example, Refs.~\onlinecite{Konno91,Meyer01}.

\section{Summary}\label{sec:summ}
We have studied the Kondo lattice model away from half filling in an external magnetic field within a mean field approach. Our mean field Hamiltonian is a generalization to high magnetic fields of the one presented in
Ref.~\onlinecite{Beach05}. This generalization is two-fold: we allow for hybridization both in the singlet and triplet channels, plus we consider the self-consistent Weiss fields in the magnetic channel. We also work in the general case in which the $c$ and $f$ electrons couple differently to the magnetic field, introducing two different gyromagnetic factors $g_c$ and $g_f$. This results in a spin-split, very massive quasiparticle Fermi surface which evolves with the magnetic field.

We showed that the self-consistent solution of this model exhibits a crossover followed by a first-order transition. At a moderate field of the order of the Kondo gap at zero field, which we called $B^*$, the spin up band enters the gap and its Fermi surface shrinks to a point, consequently disappearing from the problem. The spin down band, however, remains hybridized and mass-enhanced. This behavior is signaled by a kink in the magnetization, which changes from a nonlinear dependence for fields smaller that $B^*$, to a purely linear one after the crossover. The magnetic susceptibility shows an abrupt jump at $B^*$ and afterwards it is a constant. At a higher field $B^{\text{c}}$, approximately one order of magnitude greater than $B^*$, the heavy fermion state collapses completely. The intermediate region between $B^*$ and $B^{\text{c}}$ we have dubbed
the locked phase, since the solution of the model is (explicitly) independent of the external field. Our results were obtained in the limit of zero temperature. As ususal, it is expected that a finite temperature will have a smearing effect on the signatures of the MMT, and the kink in the magnetization will eventually disappear.~\cite{Tokiwa05} The relevant energy scale is the diference between the 
Fermi energy and the top of the lower hybridized band.

These results are in excellent qualitative agreement with the MMT shown by the heavy fermion compound  YbRh$_2$Si$_2$, which had been previously attributed to the localization of the $f$ electrons.~\cite{Tokiwa05} Quantitative agreement can also be achieved but at the expense of assuming unphysically small values for the noninteracting bandwidth $W$. Nonetheless, this does not imply any internal inconsistency in our model, since $W$ is always the biggest energy scale of the model.

The existence of the locked phase is guaranteed when we use simple estimates of the gyromagnetic factors,
based on crystal field splitting arguments. If we choose $g_f<1$, there is a very small range of the parameters $x$ and $J$ for which $B^{\text{c}}<B^*$. In this case, the heavy fermion state is destroyed before the locked state is attained and hence there is no residual heavy fermion behavior after the transition. $g_c$, in turn, controls the diamagnetic contribution to the total magnetization. We find that regime III, in which the upper band starts being filled by the spin up quasiparticles, never occurs because the heavy fermion state always collapses before reaching this regime.

\acknowledgements
S. V. K. would like to thank C. Y. Hou, A. Rahmanisisan, S. Rowley and V. M. Pereira for enlightening discussions.
\\

\appendix

\section{Mean Field decomposition} \label{sec:appA}

The mean field Hamiltonian is diagonalized by the unitary transformation 
\begin{equation*}
U = \begin{pmatrix}
U^{c+} & U^{c-}\\
U^{f+} & U^{f-}
\end{pmatrix}
= \begin{pmatrix}
\frac{-I^+}{\sqrt{(I^+)^2+V^2}} &
\frac{-I^-}{\sqrt{(I^-)^2+V^2}} \\
\frac{V}{\sqrt{(I^+)^2+V^2}} &
\frac{V}{\sqrt{(I^-)^2+V^2}}
\end{pmatrix}.
\end{equation*}
We have suppressed the wavevector and spin-projection dependence, which arises since 
$U^{an}_{\vec{k}s}$ is an explicit function of $I^n_{\vec{k}s}$ [Eq.~\eqref{Iks}].
The quasiparticles of the mean field Hamiltonian are an admixture of
the two fermion species, $
\alpha^n_{\vec{k}s} = \sum_{a=c,f}  a_{\vec{k}s} U^{a n *}_{\vec{ks}}$,
and describe the dynamics of the noninteracting system,
\begin{equation*}
\begin{split}
\hat{H}_0 
&= \sum_{\vec{k}ns}E^n_{\vec{k}s}\alpha^{n\dagger}_{\vec{k}s}\alpha^n_{\vec{k}s}
= \sum_{\vec{k}}\Bigl[ c^\dagger_{\vec{k}}
(\epsilon_{k}-\mu_c - s B_c/2)c_{\vec{k}} \\
&\qquad -
f^\dagger_{\vec{k}}(\mu_f+ s B_f/2)f_{\vec{k}}
-V_s\bigl(c^\dagger_{\vec{k}s}f_{\vec{k}s}
+f^\dagger_{\vec{k}}c_{\vec{k}}\bigr)\Bigr].
\end{split}
\end{equation*}
The complete Hamiltonian can be expressed as the sum
$\hat{H}=\hat{H}_0+\hat{H}_1$, where
\begin{equation} \label{EQ:H1explicit}
\begin{split}
\hat{H}_1 &= \sum_{\vec{k}} \biggl[
V_s\bigl(c^\dagger_{\vec{k}s}f_{\vec{k}s}
+f^\dagger_{\vec{k}s}c_{\vec{k}s}\bigr)
-\frac{1}{2}(B-B_c)c^\dagger_{\vec{k}}\sigma^3c_{\vec{k}}\\ &
-\frac{1}{2}(B-B_f)f^\dagger_{\vec{k}}\sigma^3f_{\vec{k}}
+\frac{J}{4}\sum_{\vec{q},\vec{k}'}
(c_{\vec{k}+\vec{q}}^\dagger \svec{\sigma} c_{\vec{k}})\cdot
(f_{\vec{k}'}^\dagger \svec{\sigma} f_{\vec{k}'+\vec{q}})\biggr].
\end{split}
\end{equation}
If we take expectation values of the Hamiltonian in the ground state
of $\hat{H}_0$, we arrive at a variational energy
$\mathcal{U} = \bracket{\hat{H}} = \sum_{\vec{k}ns}E_{\vec{k}s}^n\mathsf{f}^n_{\vec{k}s}
+ \bracket{\hat{H}_1}$. Here, $\mathsf{f}_{\vec{k}s}^n = \theta(-E^n_{\vec{k}s})$ is the zero-temperature
Fermi function.

In order to compute the expectation value $\bracket{\hat{H}_1}$,
we must substitute $c_{\vec{k}s} = \sum_{n=\pm} U^{c n}_{\vec{k}s}\alpha_{\vec{k}s}^n $ and $
f_{\vec{k}s} = \sum_{n=\pm} U^{f n}_{\vec{k}s}\alpha_{\vec{k}s}^n $ into the various terms in Eq.~\eqref{EQ:H1explicit}. For example,
the four-body term $(c_{\vec{k}+\vec{q}}^\dagger \svec{\sigma} c_{\vec{k}})\cdot
(f_{\vec{k}'}^\dagger \svec{\sigma} f_{\vec{k}'+\vec{q}})$ becomes
\begin{multline}\nonumber
U_{\vec{k}+\vec{q},s}^{c n\dagger}
U^{c m}_{\vec{k}r}
U_{\vec{k}'r'}^{f m'\dagger}
U_{\vec{k}'+\vec{q},s'}^{f n'}
\svec{\sigma}_{sr}\cdot\svec{\sigma}_{r's'}\\
\times \alpha_{\vec{k}+\vec{q},s}^{n\dagger} \alpha^m_{\vec{k}r}
\alpha_{\vec{k}'r'}^{m'\dagger} \alpha_{\vec{k}'+\vec{q},s'}^{n'}.
\end{multline}
Expectation values of the quasiparticle operators obey Wick's
theorem:
\begin{multline}\nonumber
\bracket{\alpha_{\vec{k}+\vec{q},s}^{n\dagger}\alpha^m_{\vec{k}r}
\alpha_{\vec{k}'r'}^{m'\dagger} \alpha_{\vec{k}'+\vec{q},s'}^{n'}}
= 
\mathsf{f}_{\vec{k}+\vec{q},s}^n(1-
\mathsf{f}^m_{\vec{k}r})\\
\times\delta^{nn'}_{ss'}\delta^{mm'}_{rr'}\delta_{\vec{k}\vec{k}'}
+
\mathsf{f}_{\vec{k}s}^n
\mathsf{f}_{\vec{k}'s'}^{n'}
\delta^{nm}_{sr}\delta^{n'm'}_{s'r'}\delta_{\vec{q}\vec{0}}
\end{multline}
Hence, using the identity $
\svec{\sigma}_{sr}\cdot\svec{\sigma}_{r's'} = 2\delta_{ss'}\delta_{rr'} - \delta_{sr}\delta_{s'r'}$ 
we find that the last term in Eq.~\eqref{EQ:H1explicit} is
\begin{equation}\nonumber
\begin{split}
\frac{1}{4}\sum_{sr}
(2-\delta_{sr})
\sum_{\vec{k}} 
U_{\vec{k}s}^{c n\dagger}
U_{\vec{k}s}^{f n}
\mathsf{f}_{\vec{k},s}^n
\sum_{\vec{k}'}
U^{c m}_{\vec{k}'r}
U_{\vec{k}'r}^{f m\dagger}
(1-\mathsf{f}^m_{\vec{k}'r})\\
+\sum_{\vec{k}ns}\frac{s}{2}
U_{\vec{k}s}^{c n\dagger}
U^{c n}_{\vec{k}s}\mathsf{f}_{\vec{k}s}^n
\sum_{\vec{k}'s'n'}\frac{s'}{2}
U_{\vec{k}'s'}^{f n'\dagger}
U_{\vec{k}'s'}^{f n'}
\mathsf{f}_{\vec{k}'s'}^{n'}.
\end{split}
\end{equation}
This result can be written compactly as
\begin{multline}\nonumber
-\frac{1}{4}\sum_{sr}
(2-\delta_{sr})
  \bracket{ c^\dagger_{\vec{k}s}f_{\vec{k}s} }
  \bracket{ f^\dagger_{\vec{k}'r}c_{\vec{k}'r} }
  +
\frac{1}{4}\bracket{c^\dagger_{\vec{k}}\sigma^3c_{\vec{k}}}
\bracket{f^\dagger_{\vec{k}}\sigma^3f_{\vec{k}}},
\end{multline}
since the term $\sum_{\vec{k}n} U^{c n}_{\vec{k}s} U^{f n \dagger}_{\vec{k}s}$ can be shown to vanish identically:
\begin{equation}\nonumber
\begin{split}
-\frac{1}{W}\sum_s \int\!d\omega\,\frac{V_s}{\omega}
&= \sum_s\frac{V_s}{W}\log\biggl(\frac{\omega_{3s}\omega_{1s}}{\omega_{4s}\omega_{2s}}\biggr)
\equiv 0.
\end{split}
\end{equation}
Finally, the expectation value of Eq.~\eqref{EQ:H1explicit} is
\begin{widetext}
\begin{equation*}
\begin{split}
\bracket{\hat{H}_1} &= \sum_{\vec{k}} \biggl[
\sum_s V_s\bracket{c^\dagger_{\vec{k}s}f_{\vec{k}s}
+f^\dagger_{\vec{k}s}c_{\vec{k}s}}
-\frac{1}{2}(B-B_c)\bracket{c^\dagger_{\vec{k}}\sigma^3c_{\vec{k}}}
-\frac{1}{2}(B-B_f)\bracket{f^\dagger_{\vec{k}}\sigma^3f_{\vec{k}}}\\
&\qquad\qquad\qquad\qquad\qquad
- \frac{J}{4}\sum_{sr\vec{k}'}
(2-\delta_{sr})
  \bracket{ c^\dagger_{\vec{k}s}f_{\vec{k}s} }
 \bracket{ f^\dagger_{\vec{k}'r}c_{\vec{k}'r} }
 +
\frac{J}{4}\bracket{c^\dagger_{\vec{k}}\sigma^3c_{\vec{k}}}
\bracket{f^\dagger_{\vec{k}'}\sigma^3f_{\vec{k}'}}\biggr].
\end{split}
\end{equation*} 
\end{widetext}

We know that ultimately the hybridization and Weiss fields are going
to behave as
$Jm_c = Q(B-B_f)$, $Jm_f = Q(B - B_c)$,
and $J\bracket{c^\dagger_{\vec{k}s}f_{\vec{k}s}}
= J\bracket{f^\dagger_{\vec{k}s}c_{\vec{k}s}} = PV_s + \bar{P}V_{-s}$,
where
$m_c = \tfrac{1}{2}\bracket{c^\dagger_{\vec{k}}\sigma^3c_{\vec{k}}}$
and
$m_f = \tfrac{1}{2}\bracket{f^\dagger_{\vec{k}}\sigma^3f_{\vec{k}}}$
and $P$, $\bar{P}$, and $Q$ are unknown factors of proportionality. The extremal values are
$P=-4/3$, $\bar{P} = 8/3$, $Q=1$, which leads to
\begin{equation*}
\bracket{\hat{H}_1} = \frac{8|V^0|^2}{3J} - \frac{8|V^3|^2}{J} 
-\frac{(B-B_c)(B-B_f)}{J}.
\end{equation*}

\section{Free Energy} \label{sec:appB}
\subsection{Conventional paramagnetic metal} 
From Eq.~\eqref{klm}, if we put the pairing channel to zero and assume $B_c=B_f=B$, we get
\begin{equation*}
\mathcal{G_{PM}} = -\frac{W}{4}(1-x^2) - \frac{g_c^2B^2}{4W} - \frac{g_f\lvert B \rvert}{2} 
+Jm_cm_f.
\end{equation*}
The corresponding magnetization is given by 
\begin{equation*}
  m_c= \frac{g_cB}{2W} \qquad \text{and}  \qquad m_f= \frac{1}{2}\sgn(B).
\end{equation*}

\subsection{Conventional itinerant ferromagnet} 
Again we put the pairing channel to zero in Eq.~\eqref{klm} but this time we keep the Weiss fields defined by $
g_cB_c = g_c B - Jm_f$ and $g_fB_f = g_f B - Jm_c $. The free energy takes the form
\begin{equation*}
\mathcal{G_{IF}} = -\frac{W}{4}(1-x^2) - \frac{g_c^2B_c^2}{4W} - \frac{g_f\lvert B_f \rvert}{2} 
-Jm_cm_f
\end{equation*}
where the magnetization is given by $m_c = \frac{g_cB_c}{2W}$ and $m_f = \frac{1}{2}\sgn(B_f)$. Hence,
\begin{equation*}
\begin{split}
\mathcal{G_{IF}} &= -\frac{W}{4}(1-x^2) - \frac{g_c^2B^2}{4W}\\&
-\frac{1}{2}\Bigl( g_f 
- \frac{g_cJ}{2W} \Bigr)B\sgn(B_f) 
- \frac{J^2}{16W} .
\end{split}
\end{equation*}

Clearly, the energy is a minimum when $
\sgn(B_f) = \sgn(B)\sgn\Bigl( 1 - \frac{g_c J}{2g_fW} \Bigr) $ 
and therefore
\begin{equation*}
\mathcal{{G_{IF}}} = -\frac{W}{4}(1-x^2) - \frac{g_c^2B^2}{4W}
-\frac{1}{2}g_f \lvert B \rvert \Bigl \lvert 1
- \frac{g_cJ}{2 g_f W} \Bigr\rvert
- \frac{J^2}{16W} 
\end{equation*}
The magnetization then is given by
\begin{equation*}
\begin{split}\nonumber
m_c & = -\frac{1}{B}\frac{\partial \mathcal{G}}{\partial g_c}
= \sgn(B)\biggl[\frac{g_c \lvert B \rvert}{2 W} - \frac{J}{4W}\zeta\biggr]\\
m_f& =  -\frac{1}{B}\frac{\partial \mathcal{G}}{\partial g_f} = \frac{1}{2}\zeta\sgn(B)
\end{split}
\end{equation*}
where $\zeta = \sgn\Bigl( 1 - \frac{g_c J}{2g_fW} \Bigr)$. Notice that when $\zeta=1$ there is a line $\lvert B \rvert = J/2g_c$ at which $m_c$
changes sign. There is no diamagnetic region when $\zeta=-1$.

The condition for the transition to be to an itinerant ferromagnet instead to a paramagnet is given by the condition $\mathcal{G}_{IF} < \mathcal{G}_{PM}$, which is satisfied for:
\begin{align*}
\lvert B \rvert &<  \frac{J}{4(g_c-1)}  & (\zeta &= 1)\\ \nonumber
\lvert B \rvert &<  \frac{J^2}{4[4g_fW - J(g_c+1)]}  & (\zeta &= -1)
\end{align*}

\vspace{5cm}
{\color{white}.}

\end{document}